\documentclass{elsarticle}
\usepackage[latin1]{inputenc}
\usepackage{amsmath}
\usepackage{amsfonts}
\usepackage{amssymb}
\usepackage{natbib}

\begin{document}
\title{Testing for Anisotropy of Space via an Extension of Special Relativity}
\author{Alon Drory}
\ead{adrory@zahav.net.il}
\address{Afeka College of Engineering, 218 Bney-Efraim Street, Tel-Aviv, 69107 , Israel}

\begin{abstract}
In special relativity, testing for spatial anisotropy usually means testing for anisotropic propagation of light. This paper explores a different possibility, in which light is still assumed to propagate isotropically in all frames with an invariant speed, yet other physical effects exhibit a direction dependence. If spatial isotropy is not assumed in the derivation of the coordinates transformations, the resulting equations differ from the Lorentz relations by an additional factor $\left( \dfrac{c - v}{c + v}\right)^{\kappa}$, where $\kappa$ is the anisotropy exponent, which depends on the direction chosen as the x-axis. Time dilation and length contractions become direction dependent. The anisotropy exponent is frame-independent, so no preferred isotropic frame exists if $\kappa$ is non-vanishing. The Doppler shift can be used to measure this exponent and determine experimentally the degree of anisotropy our universe actually possesses.
\end{abstract}


\maketitle

\section{Introduction}

The theory of special relativity is usually said to derive from two assumptions. First, the principle of relativity, which denies privileged status to any inertial frame; second, the postulate of the constancy of the speed of light, which stipulates that this velocity is independent of the velocity of the emitter and of the receiver.

Several works have tried to test the validity of these assumptions by creating alternative theories in which they are violated to some degree or other. These test theories of special relativity were naturally centered on possible violations of either the principle of the constancy of light \cite{edwards} or the principle of relativity \cite{robertson,mansouri}. 

It has long been known that the special theory of relativity is not actually based on these two assumptions alone, however \cite{levy}. One must also add space homogeneity and isotropy. These symmetries had long been held to be so fundamental as to be almost beyond need of verification. Indeed, Einstein did not even mention them as separate assumptions in his original paper \cite{einstein}, though he later admitted they were necessary.

The discovery of parity violation \cite{parity1,parity2} significantly altered this attitude. The idea that even apparently fundamental symmetries of the universe can be broken has found many uses and has gained wide acceptance. Most of these symmetries are highly formal, however, and the fundamental symmetries of space-time, homogeneity and isotropy,  remain unexamined to a large extent.

Because the theory of special relativity itself is based on a rethinking of basic concepts of space and time, it represents the natural seeting in which to explore possible violations of these fundamental symmetries. It is also a theory that remains fundamental to our description of the universe. It is therefore natural to explore whether giving up on the exactness of these symmetries appears has any measurable effect in this theory. 

This step markedly differs from other attempts at changing the basic assumptions of special relativity (SR). Initially, the principle of relativity seemed to be more obvious, because of its long history in classical mechanics, and verification of SR concentrated on the postulate of the constancy of the speed of light. It soon became clear that while the constancy of the two-way speed of light (measured along a closed path) is directly testable empirically, the one-way speed of light is another matter entirely. Measurements of this speed require a clock synchronization procedure, and the conventional character of such a procedure has given rise to an extensive literature \cite{reichenbach,grunbaum1,grunbaum2}. 

Edwards formulated a theory in which the one-way speed of light could be anisotropic, with values that depended on direction \cite{edwards}. The interpretation of such theories is delicate, however, because the time parameter that appears in the equation may not be directly related to the time experimentally measured by clocks. This happens because experimentally, one needs to synchronize clocks at different locations and the choice of synchronization method determines the relation between measured time and the time that appears in the transformations. It turns out that when this is done, Edwards's theory is empirically equivalent to standard special relativity \cite{zhang,zhang2}. In this sense, Edwards's theory is not a true test theory of special relativity.

A true test theory was formulated by Robertson \cite{robertson} and generalized by Mansouri and Sexl into the currently best known test theory of SR \cite{mansouri}. Mansouri and Sexl point out that the principle of relativity is less obvious than was previously thought because cosmological consideration seem to single out a special frame of reference, i.e., the frame in which the cosmic background radiation is isotropic. They then explore the consequences of violating the principle of relativity by assuming the existence of a single system in which synchronization by infinitesimally slow transport of clocks coincides with synchronization by light signals. That chosen frame is the only one in which the speed of light turns out to be isotropic. In all other frames, it depends on the direction of propagation. Unlike in Edwards's theory, this effect cannot be eliminated by some appropriate choice of synchronization procedure and represents a true physical difference between the preferred frame of reference and all the others.

Thus, in the Mansouri-Sexl theory, anisotropy always means anisotropy of the speed of light in some frames of reference. It has become usual to identify this effect with the very notion of spatial anisotropy in SR. The present paper takes a different view, one that is in some sense the opposite (or the complementary) of the Mansouri-Sexl one. 

In all previous modification of SR, one assumes that space-time homogeneity and isotropy hold, and tests for the influence of variations in the speed of light on the theory. This paper proceeds in the exact opposite manner. I hold on to Einstein's postulates, viz, the relativity principle and the constancy of the speed of light in all reference frames, but give up spatial homogeneity and isotropy. It must be noted that accordingly, the speed of light is assumed to be isotropic in all frames. Acording to the normal usage of the term "spatial anisotropy" in the context of special relativity, this would appear to be begging the question, since only the speed of light is ever assumed to be anisotropic in standard test theories. 

There is no logical reason why it must be so, however. Standard relativity theory endows the speed of light with unique properties. Obviously, that its value is frame-invariant does not mean that all other velocities or physical quantities are also frame-invariant, quite the reverse. The same will be shown to be the case here. The speed of light has unique properties. While its value is assumed not to depend on orientation, we will see that other properties do. Thus, I will show that the two postulates of special relativity allow for a more general form of Lorentz transformations that make length contraction and time dilation direction-dependent, even though the speed of light is not. In this sense, we are truly testing for the possible anisotropy of space, not for the anisotropy of light propagation.

The structure of the paper is as follows. Section \ref{sec:constantc} starts with the most general coordinate transformations that are compatible with the principle of relativity, assuming neither space-time homogeneity, isotropy nor the principle of constancy of the speed of light. The technical details of this somewhat lengthy but straightforward derivation appear in the appendix. These transformations are then combined with the requirement that the speed of light be identical in all reference frames. This yields automatically linear and hence homogeneous equations, which shows that adding space homogeneity to the Einstein postulates is redundant. Section \ref{sec:group} completes the derivation of the one-dimensional transformations by requiring that they form a group. Section \ref{sec:final} extends the result to three dimensional space and contains the final transformations. These turn out to differ from the Lorentz equations by a velocity dependent factor. The dependence on spatial orientation is expressed by an anisotropy exponent $\kappa$ that vanishes if space is isotropic. Its basic properties and implications are explored in section \ref{sec:kappa}. Section \ref{sec:doppler} then examines the doppler effect within such a theory in order to suggest a possible method of measuring the exponent $\kappa$. The paper concludes with a summary.

\section{Constancy of the speed of light}
\label{sec:constantc}

We seek the most general coordinate transformations from one inertial frame S to another, S', which verify Einstein's two postulates, namely the principle of relativity and the principle of the constancy of the speed of light.  

The first requirement is that inertial (uniform) motions in a frame of reference S be transformed into inertial motions in another frame S'. The most general inertiality-preserving coordinate transformations are the Fock-Lorentz transformations. The appendix contains a new derivation of this result, which makes use of more elementary mathematics than the one provided by Fock \cite{fock}. The transformations are: 
\begin{eqnarray}
\label{eq:inert-trans}
x' = \dfrac{p_1x + p_2t}{q_0 + q_1t + q_2 x} \nonumber \\
t' = \dfrac{t + Bx}{q_0 + q_1t + q_2 x}
\end{eqnarray}
where $p_1, p_2, B, q_0, q_1, q_2$ are all arbitrary constants that must now be determined from additional assumptions. 

The result Eq.(\ref{eq:inert-trans}) is completely general, in that it does \textit{not} assume either homogeneity or isotropy of space. The usual procedure is to impose these two additional conditions and then add the requirement that the velocity of light be invariant in all frames. This yields the Lorentz transformations.

I ignore space homogeneity and isotropy, and instead impose only the condition that the velocity of light be identical in all frames, independently of any properties of the light source or of the observer. Thus, in the frame S, any light signal propagating along the x-axis will be described as
\begin{equation}
\label{eq:Ssignal}
x = \pm ct + x_0
\end{equation}
where the $\pm$ sign follows from the requirement that \textit{any} light signal travel at the speed $c$, irrespective of its direction.

By inserting Eq.(\ref{eq:Ssignal}) into the general transformations, Eqs.(\ref{eq:inert-trans}), we obtain:
\begin{eqnarray}
\label{eq:translight}
x' = \dfrac{p_1x_0 + \left(p_2 \pm c p_1 \right) t}
{q_0 + \left(q_1 \pm c q_2 \right)t + q_2x_0} \nonumber \\
t' = \dfrac{Bx_0 + \left(1 \pm c B \right) t}
{q_0 + \left(q_1 \pm c q_2 \right)t + q_2x_0}
\end{eqnarray}

The principle of the constancy of the speed of light requires that the same light signal be described, in the $S'$-frame, as
\begin{equation}
\label{eq:Spsignal}
x' = \pm ct' + b_{\pm}
\end{equation}
where the constants $b_{\pm}$ may differ for each direction of propagation. 

Substituting the transformations Eqs.(\ref{eq:translight}) into Eq.(\ref{eq:Spsignal}) yields the relation:
\begin{equation}
p_1x_0 + \left(p_2 \pm c p_1 \right) t = \left(c^2 B \pm c \right) t \pm c Bx_0 +  b_{\pm}\left[q_0 + q_2x_0 + \left(q_1 \pm c q_2 \right)t \right]
\end{equation}
Comparing the time dependent and time-independent terms on both sides, we find that:
\begin{eqnarray}
\label{bpm}
b_{\pm}(q_0 + q_2x_0) = x_0 \left(p_1  \mp c B \right) \\
\label{prule}
p_2 \pm c p_1 = c^2 B + b_{\pm} \left( q_1 \pm cq_2 \right) \pm c 
\end{eqnarray}

Eq.(\ref{bpm}) shows that if $x_0 = 0$, $b_{\pm}$ also vanishes. Substituting this into Eq.(\ref{prule}), we find
\begin{equation}
p_2 - c^2 B = \pm c(1 - p_1)  \Longrightarrow \; p_2 = c^2 B \; \; ; \; \;  p_1 = 1
\end{equation}
The right hand side follows from the equations obtained by taking separately the plus and minus signs.

Returning to Eq.(\ref{prule}) and substituting the values of $p_1$ and $p_2$, we find immediately that for any $b_{\pm}$, 
\begin{equation}
b_{\pm}\left( q_1 \pm c q_2 \right) = 0  \Longrightarrow  q_1 = q_2 = 0
\end{equation}
Again, the relations follow from a comparison of the plus and minus signs in the terms.

The coordinate transformations now become
\begin{eqnarray}
\label{eq:constc}
x' = \dfrac{x +p_2 t}{q_0} \nonumber \\
t' = \dfrac{t + \left(\dfrac{p_2}{c^2}\right)x}{q_0}
\end{eqnarray}
Finally, by definition, the origin of the S'-frame, $x' = 0$, moves with respect to the S-frame at a speed $v$, which immediately implies that
\begin{equation}
p_2 = -v  \Longrightarrow B = -\dfrac{v}{c^2}
\end{equation}
and thus, 
\begin{eqnarray}
\label{eq:trans1}
x' = \dfrac{x - vt}{q_0(v)} \nonumber \\
t' = \dfrac{t - \dfrac{v}{c^2}x}{q_0(v)}
\end{eqnarray}

Note that these transformations are linear, and that this was obtained without assuming spatial homogeneity. In the standard derivation of the Lorentz equations, it is the homogeneity of space that guarantees the linearity of the equations \cite{levy}. Thus, homogeneity turns out to be redundant, and the linearity is actually guaranteed by the principle of the constancy of the speed of light on its own.

The form of the remaining parameter $q_0(v)$, which clearly may depend on $v$, is fixed by the requirement that these transformations should form a group.

\section{Group property}
\label{sec:group}
Let us introduce a third frame, S'', which moves with velocity $u'$ with respect to the S'-frame. The transformation from S' to S'' now reads
\begin{eqnarray}
\label{eq:trans2}
x'' = \dfrac{x' - u't'}{q_0(u')} \nonumber \\
t'' = \dfrac{t' - \dfrac{u'}{c^2}x'}{q_0(u')}
\end{eqnarray}

Combining these transformations with the transformations from S to S', Eqs.(\ref{eq:trans1}), we obtain the relations
\begin{eqnarray}
\label{eq:combo}
x'' = \dfrac{x - \left( \dfrac{u' + v}{1 + \dfrac{u'v}{c^2}}\right)t}{\left[ \dfrac{q_0(u')q_0(v)}{1 + \dfrac{u'v}{c^2}}\right]} \nonumber \\
t'' = \dfrac{t - \left( \dfrac{u' + v}{1 + \dfrac{u'v}{c^2}}\right)\dfrac{x}{c^2}}{\left[ \dfrac{q_0(u')q_0(v)}{1 + \dfrac{u'v}{c^2}}\right]}
\end{eqnarray}
The group property requires that this should be identical with the direct transformation from S to S'',
\begin{eqnarray}
\label{eq:direct}
x'' = \dfrac{x - ut}{q_0(u)} \nonumber \\
t'' = \dfrac{t - \dfrac{u}{c^2}x}{q_0(u)}
\end{eqnarray}
where $u$ is the velocity of the S''-frame with respect to the S frame. Comparison of this expression with the combined transformation, Eq.(\ref{eq:combo}), shows immediately that we must have
\begin{equation}
\label{eq:veloadd}
u = \dfrac{u'+ v}{1 + \dfrac{u'v}{c^2}}
\end{equation}
and the relation
\begin{equation}
\label{eq:q}
q_0 (u) = q_0\left(\dfrac{u'+ v}{1 + \dfrac{u'v}{c^2}}\right) = \dfrac{q_0(u')q_0(v)}{1 + \dfrac{u'v}{c^2}}
\end{equation}

Eq.(\ref{eq:veloadd}) is the standard relativistic velocity addition formula. Thus, this formula does \textit{not} require either spatial homogeneity or isotropy. It follows from only two assumptions, the principle of relativity and the principle of the constancy of the speed of light. No additional spatial symmetries are required.

On the other hand, $q_0$ differs from the Lorentz form. Note first that if we choose $u' = 0$ in Eq.(\ref{eq:q}), we find that 
\begin{equation}
q_0\left(v \right) = q_0(0)q_0(v) \Longrightarrow q_0(0) = 1
\end{equation}

Keeping the velocity $v$ as a constant parameter, let us define a new variable, $\epsilon$, as
\begin{equation}
\label{eq:defepsilon}
\epsilon = u - v = u' \dfrac{1 - \dfrac{v^2}{c^2}}{1 + \dfrac{u' \cdot v}{c^2}}
\end{equation}
where the rightmost side is obtained from Eq.(\ref{eq:veloadd}).

Let us define a new function $\eta(v)$ by
\begin{equation}
\label{eq:defeta}
\eta(v) = \ln{q_0(v)}
\end{equation}

Eq.(\ref{eq:q}) now becomes
\begin{equation}
\eta(v+\epsilon) = \eta(v) + \eta\left(\dfrac{\epsilon c^2}{c^2 - \epsilon v - v^2}\right) - \ln{\left[ 1 + \dfrac{\epsilon \cdot v}{c^2 - \epsilon v - v^2}\right]}
\end{equation}
Finally, we can form the derivative of $\eta(v)$ by taking
\begin{equation}
\lim_{\epsilon \rightarrow 0} \dfrac{\eta(v + \epsilon)- \eta(v)}{\epsilon} = \lim_{\epsilon \rightarrow 0} \dfrac{\eta\left(\dfrac{\epsilon c^2}{c^2 - \epsilon v - v^2}\right)}{\epsilon} - \lim_{\epsilon \rightarrow 0} \dfrac{\ln\left(1 + \dfrac{\epsilon v}{c^2 - \epsilon v - v^2}\right)}{\epsilon}
\end{equation}
Using the fact that $\eta(0) = 0$, we obtain immediately:
\begin{equation}
\label{eq:deriveta}
\dfrac{d \eta(v)}{dv} = \dfrac{1}{1 - \dfrac{v^2}{c^2}}\left[ \dfrac{d \eta}{dv}(0) - \dfrac{v}{c^2}\right]
\end{equation}
The derivative of $\eta (v)$ at $v = 0$ is a constant that we can denote as
\begin{equation}
\dfrac{d \eta }{d v}(0) = \dfrac{2}{c} \kappa
\end{equation}
where $\kappa$ is some constant. It is important to note that $\kappa$ does not depend on $v$ and is therefore frame-invariant.

Eq.(\ref{eq:deriveta}) is immediately integrable and yields:
\begin{equation}
\eta(v) = \dfrac{1}{2}\ln{\left(1 - \dfrac{v^2}{c^2}\right)} + \kappa \ln{\left( \dfrac{c + v}{c - v}\right)}
\end{equation}
From the definition of $\eta(v)$, Eq.(\ref{eq:defeta}), we finally obtain the form of $q_0$:
\begin{equation}
\label{q0}
q_0(v) = \sqrt{1 - \dfrac{v^2}{c^2}}\left( \dfrac{c + v}{c - v}\right)^{\kappa}
\end{equation}
 
\section{Final transformations}
\label{sec:final}
Next consider the transverse axes. Let a light signal be emitted in a diagonal direction in the x-y plane from the origin of the frame S at a time $t_0 = 0$. The position of a point on the wavefront is $(ct, x, y, 0)$ when observed in the S frame, and $(ct', x', y', 0)$ when observed in the S' frame. Since the speed of light is invariant, we must have that 

\begin{eqnarray}
\label{wavefront}
x^2 + y^2 = c^2t^2 \nonumber \\
x'^2 + y'^2 = c^2t'^2
\end{eqnarray}

Substituting the transformations for $x'$ and $t'$ from Eq.(\ref{eq:trans1}) and comparing with the first of Eqs.(\ref{wavefront}), we obtain

\begin{equation}
y'^2 = c^2t'^2 - x'^2 = \left(1 - \dfrac{v^2}{c^2}\right) \dfrac{y^2}{q_0^2(v)}
\end{equation}

Using Eq.(\ref{q0}) for $q_0(v)$, we obtain the final form of the transformations:
\begin{eqnarray}
\label{finalx}
x' = \left( \dfrac{1 - \beta}{1 + \beta} \right)^{\kappa} \dfrac{x - \beta c t}{\sqrt{1 - \beta^2}}  \\
\label{finaly}
y' = \left( \dfrac{1 - \beta}{1 + \beta} \right)^{\kappa} y \\
\label{finalz}
z' = \left( \dfrac{1 - \beta}{1 + \beta} \right)^{\kappa} z \\
\label{finalt}
ct' = \left( \dfrac{1 - \beta}{1 + \beta} \right)^{\kappa} \dfrac{ct - \beta x}{\sqrt{1 - \beta^2}}
\end{eqnarray}
where $\beta = \dfrac{v}{c}$, as usual. The transformation for $z'$ is obtained in the same way we obtained the transformation for $y'$.

These transformations differ from the Lorentz transformations, but one may wonder whether the difference is not merely conventional. As mentioned above, this is what happens in the Edwards theory, where a different synchronization procedure leads to apparently different transformations, but these involve a non-observable time parameter \cite{edwards}. This time parameter must then be related to observable time via the definition of a proper synchronization procedure. Once this is done, the Edwards theory turns out to be empirically equivalent to SR \cite{zhang,zhang2}. 

This is not the case here, however. The standard synchronization method is the Einstein convention, which uses light signals. Since test theories of SR alter the postulate of constancy of speed of light, this method is no longer directly usable in the new theories. This is why care must be taken to relate the time parameter that appears in the transformations to the experimental time measured by synchronized clocks. Here, on the other hand, the postulate of constancy of the speed of light remains untouched. Thus, the one-way speed of light is isotropic by definition, and the Einstein clock synchronization procedure is applicable. This means that the time obtained in the transformations is directly measurable. In Eqs.(\ref{finalx})-(\ref{finalt}), all relevant parameters  represent true physical quantities. The difference with the Lorentz transformation represents an actual physical effect, therefore, and cannot be eliminated merely by a change of conventions.

\section{Properties of $\kappa$}
\label{sec:kappa} 

The Lorentz transformations are recovered if we set $\kappa = 0$. If $\kappa$ does not vanish, however, its value will in general depend on the choice of the x-axis. 

Let S' be a frame moving at a velocity $v$ in the positive x direction with respect to the frame S. The x-axis and x'-axis are parallel. Consider an event E happening at a space-time point $\left( ct, x \right)$ in the S-frame. In the S'-frame, the position of the event is given by Eq.(\ref{finalx}).

Now let us rotate the frames by $\pi$ radians. Denote the rotated S-frame by $S_R$ and the rotated S'-frame by $S'_R$. We do not reverse the direction of motion of the frame S', and consequently,  $\beta_R = - \beta$. The event E now takes place on the negative x-axis, so that $x_R = -x$ and $x'_R = - x'$. Clearly we have that 

\begin{equation}
\label{xR1}
x'_R = \left( \dfrac{1 - \beta_R}{1 + \beta_R}\right)^{\kappa_R} \dfrac{x_R - \beta_R c t}{\sqrt{1 - \beta_R^2}}
\end{equation}

Substituting the values of $x_R$ and $\beta_R$ into Eq.(\ref{xR1}) yields now

\begin{equation}
\label{xR2}
-x'= x'_R = \left( \dfrac{1 + \beta}{1 - \beta}\right)^{\kappa_R} \dfrac{-x + \beta c t}{\sqrt{1 - \beta^2}}
\end{equation}

Comparison with Eq.(\ref{finalx}) shows immediately that 

\begin{equation}
\label{kr}
\kappa_R = - \kappa
\end{equation}

Thus, we see that if the exponent $\kappa$ does not vanish, its value depends on the orientation of the $x$-axis. This means that the exponent $\kappa$ represents a measure of the anisotropy of space.

Since $\kappa$ changes sign under a rotation of $\pi$, there must be at least one specific direction in space for which $\kappa = 0$. Along this axis, standard special relativity is valid without any alterations. Let us then denote by $\theta$ the angle of any other direction with respect to this preferred axis. Eq.(\ref{kr}) can now be rewritten as

\begin{equation}
\label{kappatheta}
\kappa(\theta + \pi) = - \kappa (\theta)
\end{equation}
which is necessarily a symmetry of the anisotropy exponent.

As one motivation for their work, Mansouri and Sexl mention that cosmological considerations hint at the special status of a specific reference frame, namely that frame in which the background cosmic radiation is isotropic. It is tempting to think of the present work in similar terms, i.e., the idea that there might be a preferred frame in which physics is isotropic, with other frames exhibiting anisotropy via the scaling factor $\left( \dfrac{1 - \beta}{1 + \beta} \right)^{\kappa}$. Such an interpretation would be incorrect, however. 

First, contrary to the Mansouri-Sexl theory, the present work assumes the validity of the principle of relativity, and hence the equal status of all reference frames. If the anisotropy of physical effects were frame-dependent, this would violate the relativity principle. Second, the anisotropy exponent, $\kappa$, was found to be independent of the velocity of the frames. Thus, it is a frame-invariant quantity. If space is anisotropic, therefore, this is not an effect that can be eliminated by transferring to some proper frame of reference. The effect will show up identically in all inertial frames.

One may think that this is an unlikely state of affairs. It would seem at first sight far more likely that there exists a cosmologically preferred frame of reference in which all directions are equivalent, with other frames exhibiting anisotropy with respect to the preferred frame, depending, e.g., on the direction of motion of the frames. 

The present analysis does not forbid such a state of affairs. Instead, it proves that if anisotropy is frame-dependent, the speed of light cannot remain frame-independent. In other words, if there exists a frame-dependent anisotropy, the speed of light will exhibit a similar frame-dependence dependency. This is basically the situation covered in standard test theories such as the Mansouri-Sexl theory. The present work seeks to investigate the more radical possibility that there exists a physical difference between directions in space, that cannot be erased by selecting an appropriate frame. However unlikely one may think this is, it is worth verifying, and Eqs.(\ref{finalx})-(\ref{finalt}) offer a numerical assessment of whether and to what extent such a preferred direction might exist.

Clearly we expect $\kappa$ to be small, otherwise we would likely have noticed its effects already. Any non-zero value of $\kappa$ will only appear in high-precision experiments, therefore. Note again that according to Eq.(\ref{eq:veloadd}), effects depending on velocity addition are isotropic, and thus cannot serve to measure $\kappa$. Since some optical measurements can be made to a very high precision, however, it would appear that this is a convenient field in which to search for the effects of spatial anisotropy. 

There is unfortunately no reliable way to extract bounds on the value of $\kappa$ from present measurements, however precise. The reason is Eq.(\ref{kappatheta}), which implies that there is at least one direction in space along which $\kappa = 0$. Thus, a single experiment will never provide reliable bounds on $\kappa$ because one cannot discard the possibility that the experiment was performed, by chance, in the direction of vanishing anisotropy exponent. For the same reason, one cannot extract a bound on $\kappa$ from existing data, since the direction in which the experiments were set up is not typically recorded. A proper measurement of $\kappa$ requires a setup in which the direction of measurement can be varied, and the results compared for different orientations. it is the variation of the value of $\kappa$ in space that would provide the signal test for the existence of spatial anisotropy.

\section{Measuring $\kappa$ through the Doppler shift}
\label{sec:doppler}

One convenient possibility of measuring $\kappa$ is through the Doppler shift. Consider a light source that emits a signal with frequency $f_0$ in its rest frame S'. Let this rest frame be moving at velocity $v$ with respect to the lab frame S and the light be emitted towards the S observer. The transformation Eq.(\ref{finalt}) implies that in the S-frame, the emitted period, $T$, differs from the rest period $T_0$ by

\begin{equation}
\label{t0}
T_0 = \left( \dfrac{1 - \beta}{1 + \beta} \right)^{\kappa} \dfrac{T - \dfrac{\beta}{c}x}{\sqrt{1 - \beta^2}}
\end{equation}

The emitter can be chosen as the origin of the frame S', so that its position in the S frame is $x = \beta c T$. Substituting this into Eq.(\ref{t0}) yields 

\begin{equation}
\label{tmove}
T = \left( \dfrac{1 + \beta}{1 - \beta} \right)^{\kappa} \dfrac{T_0}{\sqrt{1 - \beta^2}}
\end{equation}

In the S frame, the emitted wavelength is $\lambda_{emit} = c/T$, but S observes a different wavelength because the distance between two successive crests is shortened by the emitter's motion. The received wavelength $\lambda_{rec}$ is given by the standard relation 

\begin{equation}
\lambda_{obs} = \lambda_{emit} - vT
\end{equation}

Substituting the value of $T$ from Eq.(\ref{tmove}) into this relation yields

\begin{equation}
\lambda_{obs} = \left( \dfrac{1 + \beta}{1 - \beta} \right)^{\kappa - 0.5} \lambda_0
\end{equation}
where $\lambda_0 = cT_0$ is the wavelength emitted in the rest frame. Finally, the frequency that S measures is

\begin{equation}
\label{doppler}
\dfrac{c}{\lambda_{obs}} = f_{obs} = \left( \dfrac{1 + \beta}{1 - \beta} \right)^{0.5 - \kappa} f_0
\end{equation}

Thus, a precise measurement of the Doppler shift along various directions can provide a determination of the isotropy of space. Any deviation of the exponent in Eq.(\ref{doppler}) from $0.5$ would be a confirmation that space is anisotropic.

As mentioned in the previous section, much more significant would be a scan of the Doppler shift in various directions, while keeping the speed of the emitter, $v$, constant. A non-vanishing value of $\kappa$ would reveal itself as a systematic variation of the exponent in Eq.(\ref{doppler}), which must verify the symmetry relation Eq.(\ref{kappatheta}). Thus, the exponent should rise slightly as the direction of the motion of the emitter is rotated from the preferred direction along which $\kappa = 0$, then dip after completing a half-turn, finally to return to its original value once the circle is complete. A measurement of this (or a similar) effect would provide us with a quantitative estimate of the degree of anisotropy of our universe.

The transverse Doppler shift arises, as in standard relativity, from the changes in frequency only. Thus, if a source S' moves in the x-direction and emits a light signal in the y-direction, the observer in the S-frame will observe a different frequency only because of the time dilation effect, Eq.(\ref{tmove}). Hence, the transverse frequency will be observed to be 
\begin{equation}
\label{transdoppler}
f_{trans} = \left( \dfrac{1 - \beta}{1 + \beta} \right)^{\kappa}\sqrt{1 - \beta^2} f_0
\end{equation}

Because $\kappa$ depends on the orientation, identical emitters sending pulses of light transversally but along different lines should exhibit a different doppler shift, if the anisotropy exponent does not vanish.

\section{Conclusions}

In this paper, I have examined whether a possible violation of spatial isotropy and homegeneity would be detectable. Because these symmetries play a special role in the theory of special relativity, it is natural to examine the implications of their violations in this theory.

I have shown that homogeneity is recovered automatically from the postulate of the constancy of the speed of light and therefore any violation of this symmetry will not show up in the Lorentz transformations or its consequences. 

Second, the relativistic rule of addition of velocities is also independent of the assumptions of homogeneity and isotropy, and derives solely from Einstein's two original assumptions. Thus, any effect that derives from this rule will also be impervious to any violation of isotropy or homogeneity.

The coordinate transformations themselves are altered, however, if isotropy is violated. The violation appears as an additional factor that is direction dependent. The standard Lorentz transformations are now multiplied by the term $\left( \dfrac{1 - \beta}{1 + \beta}\right)^\kappa$. The exponent $\kappa$ depends on the direction of the velocity of the frame S', which is chosen to be the x-direction. The exponent changes sign upon a rotation by $\pi$ radians of the direction of the x-axis. 

The exponent is frame-independent, however, so the anisotropy under discussion here differs fundamentally from what is expected at present from cosmological considerations. The cosmological frame in which the cosmic background radiation is isotropic holds no particular position here. A non-vanishing $\kappa$ would show up experimentally even in that frame.

If such an effect does exist, I suggest that a good place to seek it is by checking the doppler shift of emitters moving in  various directions. Present data is insufficient to analyse the value of $\kappa$ because the direction in which the measurement is performed is not recorded. Since there exists at least one direction in space in which $\kappa$ vanishes, there is no way to assess whether past experiments should exhibit a measurable effect. A proper measurement of $\kappa$ requires a rotation of the apparatus along a full circle to check for possible variations in the value of the measured quantities.

Should such an effect be discovered, its implications could be profound for the description of our universe. Besides the import of the added terms in the Lorentz transformations, which would extend to every theory based on special relativity, a non-zero value of the anisotropy exponent further implies that there is a preferred direction in our universe along which the exponent vanishes, as noted above. This direction ought then to present further peculiarities of interest, which might open new fields of research in cosmology among others.

\begin{Large}\textbf{Appendix}\end{Large}
\\
\appendix

Consider two frames of reference, S and S', the latter moving with respect to the former with a velocity $v$ directed along the positive $x$ axis. We seek the transformations of the space time coordinates from one frame to another, assuming only that both frames are inertial, i.e., we seek the explicit form of the functions:
\begin{eqnarray}
\label{eq:x-t}
x' = F(x,t) \\
t' = G(x,t)
\end{eqnarray}

That the two frames are inertial means that the law of inertia holds in both. We assume, arbitrarily, that S is inertial, so that isolated bodies move uniformly with respect to this frame. For S' to be inertial requires that such bodies also move uniformly with respect to S'. Hence, if $x=ut+x_{0}$ (where $u$ is the velocity in the frame S and $x_{0}$ the initial position), then the position of the same body when referred to the frame S' must be of the form: 
\begin{displaymath}
\label{eq:inertial}
x' = a(u,x_{0})t' + b(u,x_{0}) 
\end{displaymath}
where $a(u,x_{0})$ is the body's velocity in S' and $ b(u,x_{0})$ its initial position in S'. Both quantities obviously depend on $u$ and $x_{0}$, the corresponding quantities in the frame S. Substituting these expressions into Eq.(\ref{eq:x-t}) yields the relation:
\begin{displaymath}
\label{eq:F-G}
a(u,x_{0})G(ut + x_{0},t) + b(u,x_{0}) = F(ut + x_{0},t)
\end{displaymath}

Since this must hold for all values of $u$ and $x_{0}$, we can set $u=0$. Furthermore, since $x_{0}$ is arbitrary, we can rewrite it simply as $x$. Hence, Eq.(\ref{eq:F-G}) now becomes
\begin{displaymath}
a(0,x)G(x,t) + b(0,x) = F(x,t)
\end{displaymath}
thus yielding a general relation between $F(x,t)$ and $G(x,t)$ that is valid for all $x$ and $t$.
Let us define two functions of a single variable:
\begin{eqnarray}
\label{eq:def-ab}
\alpha(x)= a(0,x) \\
\beta(x)= b(0,x)
\end{eqnarray}
We can now write our transformation as:
\begin{eqnarray}
\label{eq:transgen}
x'(t) = \alpha(x)G(x,t)+\beta(x) \\
t' = G(x,t)
\end{eqnarray}
where $G(x,t), \alpha(x)$ and $\beta(x)$ remain to be determined. Substituting $x = ut + x_{0}$ and $x' = a(u,x_{0})t'+ b(u,x_{0})$ again, we obtain, after rearranging the terms slightly:
\begin{equation}
\label{eq:b1}
b(u,x_0) = \left[\alpha(ut + x_0) - a(u,x_0) \right] G(ut + x_0, t) + \beta(ut + x_0)
\end{equation}
Now the left hand side must be independent of the time t. The functional forms of $G(x,t), \alpha(x), \beta(x)$ and $a(u,x_0)$ are then completely determined by the requirement that the right hand side must also be independent of t. 

To begin with we can eliminate $b(u,x_0)$ by setting t = 0 in the right hand side. This yields the relation:
\begin{equation}
\label{eq:b2}
b(u,x_0) = \left[ \alpha(x_0) - a(u,x_0) \right] G(x_0, 0) + \beta(x_0)
\end{equation}
Subtracting this from Eq.(\ref{eq:b1}), we obtain the relation
\begin{eqnarray}
\label{eq:basic}
0 = \left[ \alpha(ut + x_0) - a(u,x_0)\right]G(ut + x_0,t) + \beta(ut+x_0) - \beta(x_0) \nonumber \\
- \left[\alpha(x_0) - a(u,x_0) \right] G(x_0, 0) 
\end{eqnarray}
This is the fundamental relation from which we can determine the form of all the unknown functions.

For simplicity we can further assume that the two frames we compare have been set up so that the time $t = 0$ is the moment when both origins coincide, and that this is also the moment from which the time $t'$ is measured. This implies that at the time $t = 0$, we have:
\begin{eqnarray}
\label{choice1}
0 = t' = G(x,t)\vert_{x=0, t=0} = G(0,0)\\
0 = x'(t'=0) = \left[\alpha(0) + a(0,0) \right]G(0,0)+\beta(0) \Longrightarrow  \beta(0) = 0
\end{eqnarray}

Next, set $x_0 = 0$ in Eq.(\ref{eq:basic}). Furthermore, since the parameter $u$ is arbitrary, we can define, for any pair $x,t$, that $u = \frac{x}{t}$. Replacing the parameter $u$ by this ratio and using Eq.(\ref{choice1}), we can isolate the function $G(x,t)$ and express it as:
\begin{equation}
\label{eq:G1}
G(x,t)= \frac{\beta(x)}{a\left(\dfrac{x}{t} , 0 \right) - \alpha(x)}
\end{equation}
This reduces the number of unknown functions by one, and leaves us with $\alpha(x) , \beta(x)$ and $a(u,0)$ still to determine. This is done by noting that the basic relation, Eq.(\ref{eq:basic}), implies that the right hand side must be independent of time, so that all its derivatives must vanish identically. Retaining only the elements that depend on time, we find that:
\begin{eqnarray}
\label{eq:deriv}
\frac{d}{dt}\left\lbrace\left[ \alpha(ut + x_0) - a(u,x_0)\right]G(ut + x_0,t) + \beta(ut+x_0)\right\rbrace= 0 \\
\label{eq:deriv2}
\frac{d^2}{dt^2}\left\lbrace\left[ \alpha(ut + x_0) - a(u,x_0)\right]G(ut + x_0,t) + \beta(ut+x_0)\right\rbrace = 0 
\end{eqnarray}
In the first derivative, set $x_0= 0$ and $t = 0$. We can now isolate the function $a(u,0)$ and obtain:
\begin{equation}
\label{eq:a-func}
a(u,0) = \alpha(0) + \frac{A \cdot u}{ 1 + B \cdot u}
\end{equation}
where the constants $A, B$ are defined as:
\begin{eqnarray}
\label{eq:defAB}
A = \dfrac{\left(\dfrac{d \beta(0)}{dx}\right)}{\left(\dfrac{\partial G(0,0)}{\partial t}\right)} \\
B = \dfrac{\left(\dfrac{\partial G(0,0)}{\partial x}\right)}{\left(\dfrac{\partial G(0,0)}{\partial t}\right)} 
\end{eqnarray}
subsituting the result Eq.(\ref{eq:a-func}) for $a(u,0)$ back into the expression for $G(x,t)$, Eq.(\ref{eq:G1}), we obtain for this function the form:
\begin{equation}
\label{eq:G2}
G(x,t) = \frac{\beta(x)\left( t + Bx\right)}{Ax - \delta(x)\left(t + Bx\right)}
\end{equation}
where we defined the function:
\begin{equation}
\label{eq:defdelta}
\delta (x) = \alpha(x) - \alpha (0)
\end{equation}
Finally, consider the second derivative in Eq.(\ref{eq:deriv2}). After calculating the derivative, we set $t = 0$ and rewrite $x_0$ as $x$, since this is an arbitrary parameter. This yields the expression:
\begin{eqnarray}
\label{eq:2deriv}
0 = u^2\frac{d^2 \delta(x)}{dx^2} G(x,0) + 2u \frac{d \delta (x)}{dx}
\left[u \frac{\partial G(x,0)}{\partial x} + \frac{\partial G(x,0)}{\partial t} \right] \nonumber \\
- \left\lbrace \frac{u \frac{d \beta(x)}{dx} + \frac{d \delta (x)}{dx} G(x,0)}{u \frac{\partial G(x,0)}{\partial x} + \frac{\partial G(x,0)}{\partial t}}\right\rbrace \left[u^2 \frac{\partial^2 G(x,0)}{\partial x^2} + 2u \frac{\partial^2 G(x,0)}{\partial x \partial t} + \frac{\partial^2 G(x,0)}{\partial t^2} \right]
\end{eqnarray}
This rather complex expression can now be simplified. We divide by $u$ and multiply by $u\dfrac{\partial G(x,0)}{\partial x} + \dfrac{\partial G(x,0)}{\partial t}$. The parameter $u$ doesn't appear in any of the functions $G(x,0), \delta (x), \beta(x)$ or their derivatives, so all the powers of $u$ appear explicitly as coefficients. Collecting the various terms according to their powers of $u$, we end up with a quadratic equation:
\begin{displaymath}
\label{eq:quad}
0 = C_2u^2 + C_1 u + C_0
\end{displaymath}
where the coefficients $C_i$ are given by:
\begin{eqnarray}
\label{eq:coeffic}
C_0 = 2 \frac{d \delta (x)}{dx}\left[\frac{\partial G(x,0)}{\partial t}\right]^2 - \left[\frac{d \beta (x)}{dx} + \frac{d \delta (x)}{dx} G(x,0) \right]\frac{\partial^2 G(x,0)}{\partial t^2}\\
C_1 = \frac{\partial G(x,0)}{\partial t}\frac{d^2 \delta (x)}{dx^2}G(x,0) + 4 \frac{d \delta(x)}{dx}\frac{\partial G(x,0)}{\partial x}\frac{\partial G(x,0)}{\partial t} \nonumber \\
- 2 \frac{\partial^2 G(x,0)}{\partial x \partial t}\left[\frac{d \beta (x)}{dx} + \frac{d \delta (x)}{dx} G(x,0) \right] + \frac{\partial G(x,0)}{\partial t} \frac{d^2 \beta(x)}{dx^2}\\
C_2 = \frac{\partial G(x,0)}{\partial x}\frac{d^2 \delta (x)}{dx^2}G(x,0) + 2 \frac{d \delta(x)}{dx}\left[\frac{\partial G(x,0)}{\partial x}\right]^2 \nonumber \\
- 2 \frac{\partial^2 G(x,0)}{\partial x^2} \left[\frac{d \beta (x)}{dx} + \frac{d \delta (x)}{dx} G(x,0) \right] + \frac{\partial G(x,0)}{\partial x} \frac{d^2 \beta(x)}{dx^2}
\end{eqnarray}
Clearly, every one of these coefficients must vanish separately.
Consider first the coefficient $C_0$. Using the form of $G(x,t)$ given by Eq.(\ref{eq:G2}), we end up with the following expression for this coefficient:
\begin{displaymath}
\label{eq:a0}
C_0 = \frac{2Ax \beta(x)}{\left[A x - x\delta (x)B\right]^3} \left[\beta(x) \frac{d \delta (x)}{dx} - \delta (x) \frac{d\beta (x)}{dx} \right] = 0
\end{displaymath}
The possibility $A = 0$ yields that $a(u,x_0) = \alpha(0)$, i.e., the velocity measured in the primed frame would be independent of the velocity $u$ in the unprimed frame, which is clearly impossible. Similarly, the possibility that $\beta (x) = 0$ leads to $G(x,t) = 0$ for any $x, t$ which is again meaningless. The vanishing of $C_0$ must therefore imply that:
\begin{displaymath}
\beta(x) \frac{d \delta (x)}{dx} - \delta (x) \frac{d\beta (x)}{dx} = 0
\end{displaymath}
The solution of this equation is 
\begin{equation}
\label{eq:beta-delta}
\delta (x) = D \beta(x)
\end{equation}
where $D$ is some constant.

The vanishing of $C_1$ and $C_2$ turns out to yield identical results, so we can look at, e.g., $C_1$ to determine the single remaining unknown function $\beta (x)$. Using the previous results Eqs(\ref{eq:G2}) and (\ref{eq:beta-delta}), we obtain:
\begin{displaymath}
0 = C_1 = \frac{A^2}{x\left[ A - B D \beta(x) \right]^3} \Bigg\lbrace \beta(x) \frac{d^2 \beta (x)}{dx} - \frac{2}{x}\left[x \frac{d \beta(x)}{dx} - \beta(x) \right] \frac{d\beta(x)}{dx} \Bigg\rbrace
\end{displaymath}
The expression in the curly brackets must vanish, which can be shown to imply that:
\begin{displaymath}
\frac{d}{dx} \ln{\left[\frac{x^2 \frac{d\beta(x)}{dx}}{\beta^2 (x)}\right]} = 0
\end{displaymath}
This is immediately integrable to yield:
\begin{displaymath}
\label{eq:beta}
\beta (x) = \frac{x}{p \cdot x} + q
\end{displaymath}
where $p, q$ are arbitrary constants. 

Substituting this back into the expression for $G(x,t)$, Eq.(\ref{eq:G2}), and using the expression for $\delta(x)$, Eq.(\ref{eq:beta-delta}), we find finally:
\begin{displaymath}
\label{eq:finalG}
G(x,t) = \frac{t + Bx}{q_0 + q_1t + q_0 x}
\end{displaymath}
where we define:
\begin{eqnarray}
\label{eq:defq}
q_0 = q \cdot A \\
q_1 = -D \\
q_2 = p \cdot A - B \cdot D
\end{eqnarray}
Of course, since $A,B,D,p,q$ are all arbitrary, the specific definitions of the $q_i$ do not matter. These are merely an alternative set of arbitrary constants.

We can now substitute this expression back into Eq.(\ref{eq:transgen}) to obtain the result for $x'$. The final form of the transformation can now be written as:
\begin{eqnarray}
x' = \frac{p_1x + p_2t}{q_0 + q_1t + q_0 x} \nonumber \\
t' = \frac{t + Bx}{q_0 + q_1t + q_0 x}
\end{eqnarray}
where $p_1, p_2, B, q_0, q_1, q_2$ are all arbitrary constants.

\end{document}